\documentstyle[mystyle,graphicx,epsfig]{article}
\newcommand{\be}{\begin{eqnarray}}
\newcommand{\ee}{\end{eqnarray}}
\renewcommand{\indent}{\mbox{}\hskip 0.5cm}
\begin{document}
\title{Development and Evolution of Neural Networks in an Artificial Chemistry}
\author{Jens C.~Astor\thanks{$^\star$On leave from 
Dept.~of Medical Informatics, University
    of Heidelberg/School of Technology, Heilbronn, Germany, 
jastor@stud.fh-heilbronn.de}\and
Christoph Adami\\\mbox{}\\ Computation and Neural Systems and 
        Kellogg Radiation Laboratory \\
        California Institute of Technology\\ 
        Pasadena, CA, 91125 (USA)}
\maketitle
\vskip 1cm 
\begin{abstract}
  We present a model of decentralized growth for
  \textit{Artificial Neural Networks} (ANNs) inspired by 
  the development and the physiology of real nervous systems.  In
  this model, each individual artificial neuron is an autonomous unit
  whose behavior is determined only by the {\em genetic information} it
  harbors and {\em local concentrations} of substrates modeled by a 
  simple artificial chemistry. Gene expression is manifested as axon and
  dendrite growth, cell division and differentiation, substrate 
  production and cell stimulation. We demonstrate the model's power with a
  hand-written genome that leads to the growth of a simple network which 
  performs classical conditioning. To evolve more complex structures, we
  implemented a platform-independent, asynchronous, distributed
  \textit{Genetic Algorithm} (GA) that
  allows users to participate in evolutionary
  experiments via the {\em World Wide Web}.
\end{abstract}

\section{Introduction}

Ever since computational neuroscience was born with the introduction
of the abstract neuron by McCulloch and Pitts in
1943~\cite{culpitts1}, we have witnessed a gap between the
mathematical modeling of neurons---inspired by Turing's notions of
universal computation---and the physiology of biological neurons and
the networks they form. The current state of affairs reflects this
dichotomy: neurophysiological simulation test beds~\cite{bower1}
cannot solve engineering problems, while sophisticated ANN
models~\cite{fahlleb1} do not
explain the miracle of biological information processing.

Compared to real nervous systems, classical ANN models have a
serious shortcoming owing to the fact that they are engineered to
solve particular classification problems, and analyzed according to
standard theory based mainly on statistics and global error reduction.
As such, they can hardly be considered universal.  Hence, such models
{\em define} the network architecture {\em a priori} which is in most
cases a fixed structure of homogeneous computation units.\\
\indent Some models support problem-dependent network changes during
simulation~\cite{fritzke1,fahlleb1}. In these models, global decisions
lead to network structures adapted to the problem at hand.  Other
approaches try to shape networks for a particular problem by evolving
ANNs either directly~\cite{Ackley92}, or indirectly via a growth
process~\cite{gruau1}. More recently, approaches like
Ref.~\cite{michel1} include a kind of artificial chemistry which
allows a more natural development. Still, in these models neurons are
unevolvable homogeneous structures in a more or less fixed
architecture which, we believe, limits their relevance to natural
nervous systems.

In this paper we investigate the idea that interesting
information-processing structures can be grown from a model which
follows four basic principles of molecular and evolutionary biology,
listed below. While models for ANNs currently exist that implement a
selection of them, the inclusion of all four opens the 
possibility that, given enough evolutionary time, novel structures can
emerge that are comparable to natural 
nervous systems.

\begin{itemize}
\item{\sc Coding}. The model should encode ANNs in such way that 
evolutionary principles can be applied.
\item{\sc Development}. The model should be capable of growing an ANN by
    a completely decentralized developmental process, based {\em
    exclusively} on the cell and its interactions.
\item{\sc Locality}. Each neuron must act autonomously and be
    determined only by its genetic code and the state of its {\em
    local} environment.
\item{\sc Heterogeneity}. The model must have the capability to describe
    different, {\em heterogeneous} neurons in the same ANN.
\end{itemize} 

One of the key features of a model implementing those principles will
be the absence of explicit activity functions, learning rules, or
connection structures. Rather, such characteristics should emerge
in the adaptive process and lead to ANNs with open architectures and
more universal artificial neurogenesis.

While keeping in mind that the model is not designed
to reproduce real neural systems, we posit that an adherence to the
fundamental tenets of molecular and evolutionary principles---albeit
in an artificial medium---represents the most promising unexplored avenue in
the search for intelligent information-processing structures.

\section{Model}
In this section we introduce our model of neurogenesis starting with
the artificial physics and biochemistry, and go on to explain how
local gene expression ultimately results in information-processing
structures. This gene expression takes place exclusively in artificial
neurons which are embedded in a tissue-like structure. As this model
is inspired by the concepts of molecular cell
biology, we use the nomenclature of this science unabashedly while
issuing the caveat that they are analogical in nature only.

\subsection{Artificial Physics}

The physical world is a two-dimensional grid of hexagons.
Each such site harbors certain concentrations of
substrates, measured as percentage values of saturation between 0 and
1. As cells are equidistant in a hexagonal lattice, the diffusion of substrate
$k$ in cell $i$ can be modeled 
discretely as
\be
C_{ik}(t+1)=\frac{D}{6}\sum_{j=1}^{6}\left(C_{ik}(t)-C_{N_{i,j}k}(t)
\right)
\ee
where $C_{ik}(t)$ is the concentration of substrate $k$ in site $i$,
$D$ is a diffusion coefficient ($D<0.5$ to avoid
substrate oscillation), and $N_{i,j}$ represents the $j$th neighbor
of grid element $i$.

\begin{figure}[h]
\centering
\includegraphics[angle=-90, width=8 cm]{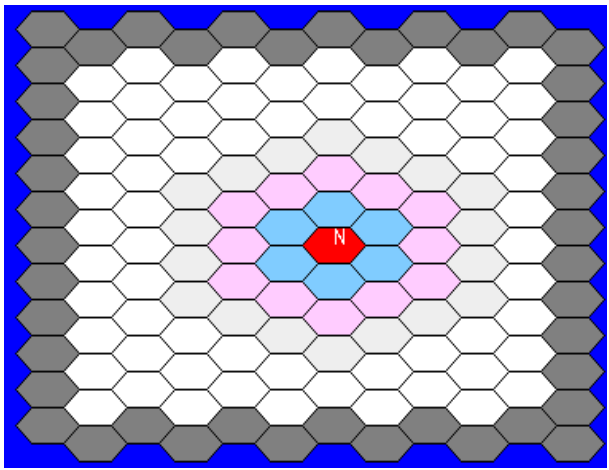}
\vskip 0.5cm 
\caption{{\small Hexagonal grid with boundary elements. Diffusion
  occurs from a local concentration peak at grid element $N$.}}
\label{fig:difffig}
\end{figure}

Accordingly, a local concentration of substrate will diffuse through the 
tissue under conservation of mass. The tissue itself is surrounded by
special boundary elements which absorb substrates (Figure
\ref{fig:difffig}), thus modeling diffusion in infinite space. 
We caution at this point that the hexagons are sites that may {\em
  harbor} cells, but otherwise only represent a convenient
equidistant discretization of space to facilitate the distribution of
chemicals via diffusion.

\subsection{Artificial Biochemistry}
We distinguish four different classes of substrates:
\begin{itemize}
\item{\sc External proteins}: Diffusive substrates which can be
  produced by neurons if expressed.
\item{\sc Internal proteins} : Produced by neurons, but non-diffusive
as they cannot cross cell membranes.
\item{\sc Cell-type proteins}: Each neural cell harbors an external
 protein that defines its type. Like any external protein it is diffusive.
\item{\sc Neurotransmitter}: Special type of internal protein used for
  directed information exchange between neurons. 
\end{itemize} 

\subsection{Artificial Cell}

{\bf Cell types}\quad 
We distinguish three kinds of neurons on the cellular level:
actuator cells, sensor cells, and common neurons. The first two types
are special versions  of the third and are used as interface to a
(simulated) environment to which the network adapts and on which it
computes. These neurons can be excited to a real-valued level
between 0 and 1, and take part in the information transfer via
dendritic or axonal connections, respectively (see below). 
Each type of cell is also characterized by its own cell-type
protein, which it produces continuously at a certain rate.
These cell-type proteins diffuse over the tissue (Figure
\ref{fig:difffig}) and therefore signal cell existence to other
cells. They can be compared to {\em growth factors} known from the development
 of real nervous systems.

Actuator and sensor cells do not carry genetic information; they are
used solely as interfaces to the environment (input-output units, see
Section \ref{sec_simulation}.) They represent {\em sources} and {\em
  sinks} of signal. Consequently, their behavior is hardwired and 
does not depend on transcription as for common neurons.  
The latter can receive a flux of neurotransmitter from dendrites 
with a particular weight. However,
this does not imply an automatic stimulation of activity unless such
behavior is explicitly encoded in the neuron's genome. Table
\ref{tab:CellTypes} summarizes the cell types and how they interact
with other computational elements used in our model.

\begin{figure}[h]
\includegraphics[angle=-90, width=8.5cm]{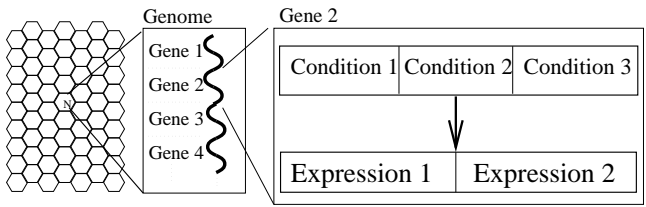}
\vskip 0.5cm 
\caption{Genetic structure of neural cells. Local
  concentrations of substrates (condition atoms) trigger gene
  expression.}
\label{fig:Genome}
\end{figure}

\begin{table}
\begin{tabular}{|l|c|c|c|c|c|c|}\hline
\em Type &\em [1] &\em [2] &\em [3] &\em [4] &\em [5] &\em [6] \\\hline
Neuron &x&x&x&x&x&x\\
Sensor  &x&x&&x&&x\\
Actuator &x&x&&&x&x\\
Grid element &x&&&&&\\
Boundary element &&&&&&\\
\hline
\end{tabular}
\vskip 0.5cm 
\caption{Features of different cell types building the
  artificial organic tissue:
[1] participates in diffusion,
[2] can be stimulated,
[3] depends on gene expression,
[4] can have axons,
[5] can have dendrites,
[6] produces a diffusible cell-type protein.}
\label{tab:CellTypes}
\end{table}
%
\noindent {\bf Genetic code and gene expression in artificial
  neurons}\quad
\label{sec_geneexpression}
Each neural cell carries a genome which completely encodes its
behavior.  Genomes consist of genes which can be viewed as a genetic
program that can either be executed ({\em expressed}) or not,
depending on a gene {\em condition} (akin to the regulator/operator genes in the Jacob-Monod-model). A gene condition is a combination
of several condition {\em atoms}, usually related to local
concentrations of substrates. The expression of a gene 
can result in different behaviors such as the production of a protein, cell
division, axon/dendrite growth, cell stimulation, etc. Figure
\ref{fig:Genome} clarifies the structure of the genetic code. Thus,
gene conditions model the influence of external concentrations on the
expression level of the gene, i.e., they model activation and suppression
sites. 
To evaluate a gene condition, each element of its chain of
condition atoms contributes to obtain a real-valued  expression level
between 0 and 1, describing the strength with which the respective
gene expression will take place. 
\begin{table}
\begin{tabular}{|l|l|}\hline
\em Condition &\em Description\\\hline
\texttt{ADD [EP]} &$\Phi_{\rm new}$=$\Phi_{\rm before}+\texttt{[EP]}$\\
\texttt{MUL [eNT]} &$\Phi_{\rm new}$=$\Phi_{\rm before}\times\texttt{[eNT]}$\\
\texttt{SUP [CTP0]} &suppresses gene if cell not of
type \texttt{CTP0}\\
\texttt{AND [IP]} &fuzzy AND: 
$\Phi_{\rm new}=\max(\texttt{[IP]},\Phi_{\rm before})$\\
\hline
\end{tabular}
\vskip 0.5cm
\caption{Condition atoms and their interpretation. Condition
  atoms build a gene condition, which is obtained by
  evaluating its condition atoms in the  given order. Here,
  \texttt{[XY]} means `the current local concentration of substrate
  \texttt{XY} inside of the cell to which the gene condition belongs'.
  $\Phi$ is the evaluation result of this gene condition.}
\label{tab:Cond}
\end{table}

\begin{figure}[!tbp]
\centering
\epsfig{file=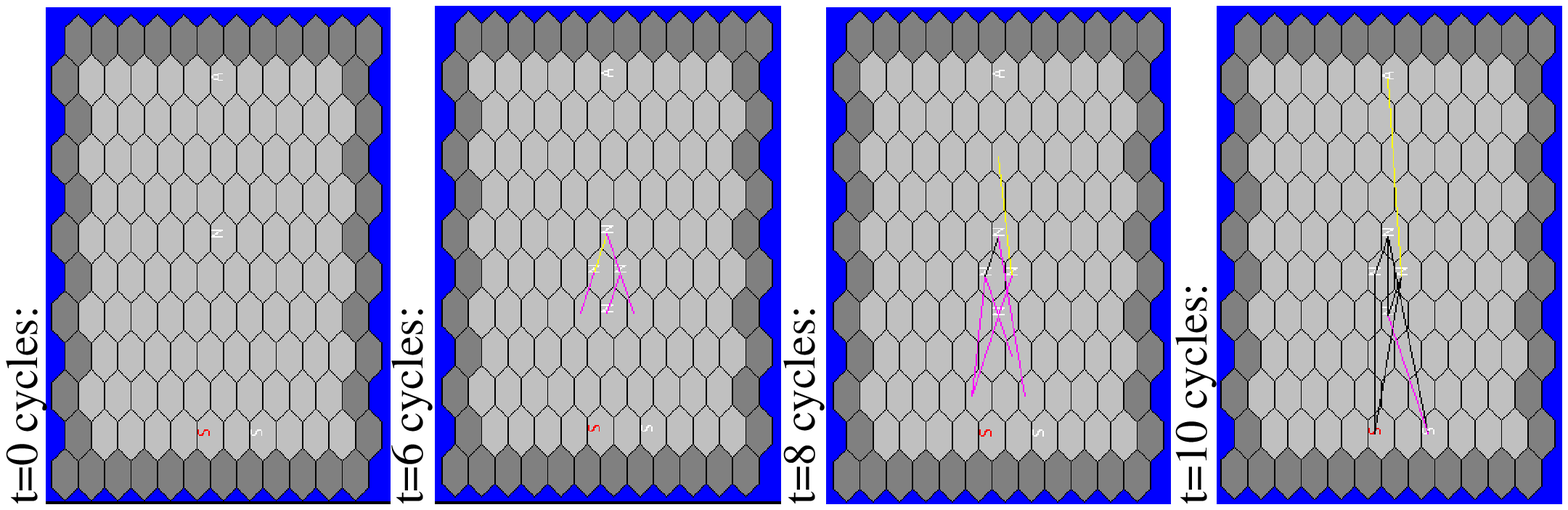, angle=-90, width=7.0cm}
\caption{{\small \label{fig:bedRefl3}Development of the network for 
classical conditioning.}}
\end{figure}

Consider for example substrates with local concentrations
\small\texttt{[ep0, ip, ep1]=(0.3, 0.5, 0.5}). \normalsize Then, the
\begin{table*}[!]
\centering
\begin{tabular}{|l|l|l|}\hline
\em Expression & \em Command Description& \em Influence of\\
&\em &\em Condition Value\\\hline
\texttt{PRD[XY]} &produce substrate \texttt{XY}& production quantity\\
\texttt{GDR[XY]} &grow dendrite following gradient of \texttt{XY}& growing probability\\
\texttt{GRA[XY]} &grow axon following gradient of \texttt{XY}&growing probability\\ 
\texttt{SPL[CTPx]} &divide to \texttt{CTPx}-type cell&probability\\
\texttt{EXT} &excitatory stimulus & increase rate\\
\texttt{INH} &inhibitory stimulus & decrease rate\\
\texttt{MOD+} &increase connection weights & strengthening factor\\
\texttt{MOD-} &decrease connection weights & weakening factor\\
\hline
\end{tabular}
\vskip 0.5cm 
\caption{Overview of expression commands. Growing
  axons/dendrites follow the substrate gradient until the local
  maximum is reached, then connect to the cell (if it exists). 
  Strengthening/weakening is a percentage increase/decrease
  of connection weights, determined by the product of the last
  neurotransmitter influx at each connection and the value of the gene
  condition. The cell-type protein assigned to a cell division
  command determines the future type of the offspring cell.
  In this example, the daughter will be of type \texttt{CTPx} and
  therefore produce cell-type protein \texttt{CTPx} 
continuously.}
\label{tab:Expr}
\end{table*}
evaluation of gene condition \small\texttt{ADD[ep0] ADD[ip] MUL[ep1]}
\normalsize would lead to an overall expression level \small 0.4
\normalsize (this value is modded back into range between 0 and 1 if
it falls outside it).  Table \ref{tab:Cond} illustrates a few examples
of such conditions, while
Table \ref{tab:Expr} gives an overview about the different
gene {\em expression} commands. The evaluation result of the
gene condition has different meanings depending on the
gene expression command.

\subsection{Simulation of the Artificial Organism}
\label{sec_simulation}

\begin{figure*}[!tbp]
\begin{verbatim}
1. NNY(ip) SUP(cpt) ANY(spt0) -> SPL(acpt0) PRD(ip) SPL(acpt2) GDR(spt0) DFN(NT1)
2. NNY(ip) SUP(acpt0) ANY(spt1) ANY(cpt) -> PRD(ip) GDR(spt1) GRA(cpt) DFN(NT1)
3. ANY(spt1) SUP(acpt2) NNY(ip) ANY(apt0) -> SPL(acpt1) GDR(spt1) PRD(ip) GRA(apt0)
4. ANY(acpt2) SUP(acpt1) ANY(spt0) NNY(ip) -> GRA(acpt2) GDR(spt0) GDR(cpt) PRD(ip)
5. ANY(ip) -> PRD(ip)
6. NSUP(cpt) NSUP(acpt1) ADD(eNT) -> EXT
7. SUP(acpt1) ADD(NT1) MUL(eNT) -> EXT
8. ADD(eNT) -> PRD(ip1)
9. ADD(ip1) -> PRD(ip2)
10. SUP(cpt) ADD(NT1) MUL(ip2) -> PRD(ep)
11. SUP(cpt) ADD(ep) -> EXT
\end{verbatim}
\caption{Genome for development and behavior of network exhibiting
  classical conditioning \label{genome}} 
\end{figure*}

The tissue of cells produced by gene expression and cell growth is
termed the {\em artificial organism}. It receives input from the
environment (the outside world) and can act on it by signaling to the
environment via its actuators. In the simplest case, thus, the
organism receives and generates patterns.

A simulation always starts by creating sensor and actuator
cells. Their number is determined only by the complexity of the outside world
and is not coded for in the genome. In other words, these cells
really represent {\em possible} signals and actuations in the world,
not actual signals and actuations performed by the organism. An
organism chooses to receive input or perform an actuation by
connecting to these cells. 
If needed, an additional reinforcement cell can be
created. This is a special sensor cell (with its own
cell-type protein) used to provide a reinforcement signal
from the world about the behavior of the organism. Whether or not
this signal is used is determined by the organism's genome. Furthermore, at the
start of each simulation, one
initial neuron is placed in the center of the grid.
After initialization, the simulation can begin. Input from the
world is provided to the sensor cells, diffusion of produced
cell-type proteins and external proteins takes place,
and neurons execute their genetic code synchronously.

Depending on its gene expression, a neuron starts
growing axons and dendrites, produces offspring
cells and might initiate cell
differentiation. Gene expressions may lead to protein
production cascades, stimulation, and ultimately information exchange
between neurons.
After every simulation cycle the network's `fitness' is determined by
comparing any inputs and outputs to what is expected in this
particular world, producing a real-valued reinforcement signal between 0
(punishment) and 1 (reward). This signal can be used by the organism  if a
reinforcement sensor is present and if the organism chooses to connect
to it.

\section{An Example Genome}
\begin{figure}[h]
\begin{center}
\epsfig{file=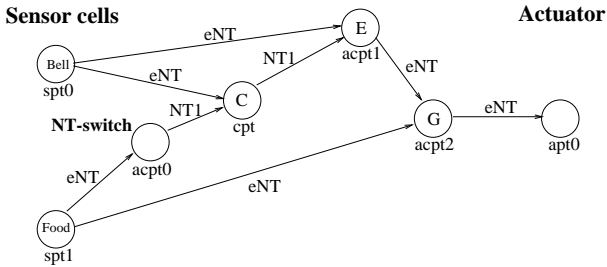, angle=-90,width=8 cm}
\caption{{\small \label{fig:schematicBedRefl} A schematical
    representation of the network for classical conditioning. The
    types of neurotransmitter used are shown next to the axons. The
    cell-type protein used by each cell is indicated near the cell body.}}
\end{center}
\end{figure}

Figure \ref{fig:bedRefl3} documents the
development of a simple ANN from a hand-written genome. 
Starting from a single initial
neuron, cell division takes place and connections
(axons and dendrites) start to grow along
the gradient of diffusing cell-type proteins. After a
while, sensors, neurons and actuator cells
are connected in a particular manner. In fact, this network displays
conditioned reflex behavior as in Pavlov's classical
experiment~\cite{pavlov}.
Suppose the sensor on the lower left side in Figure \ref{fig:bedRefl3}
is stimulated (active) at the sound of a bell. Further, suppose
the upper left sensor is an {\em optical stimulus} representing the 
presence (or absence) of food.
Finally, let us imagine that the actuator on the right side triggers a
salivary gland if food is present. This behavior is the hardwired
unconditioned reflex. The above network can learn to associate the
reflex with a condition: the sound of the bell. If presence of food 
and the ringing of the bell are associated repeatedly, the network will
learn to trigger the gland even if {\em only} the bell rings. 
If the bell rings after the conditioning without the presence of food,
the association will gradu\-ally, but steadily, weaken.
Such a behavior can be modelled using different kinds of cell-types. 
The ``C'' cell is activated if the network is in the conditioned
state, which means that acoustical and optical stimuli have been
present together before. Cell ``E'' is activated if the acoustical
stimulus is currently present and the network is in the conditioned
state at the same time. If so, cell ``G'' representing the trigger of
the salivary gland is activated. Of course, cell ``G'' is also
activated if only food is present. This is the ``hardwired'' reflex. A
schematical drawing of the network is shown in 
Figure~\ref{fig:schematicBedRefl}. 
The genome which encodes the development and behavior of this
network is shown in Fig.~\ref{genome}.

\indent It is beyond the scope of this paper to
go into the details of this ge\-nome and its function (see
\cite{astorthesis,astorpaper} for a more thorough description). However, the explanation
that follows still gives an idea of
the type of information necessary to grow networks with particular
characteristics.

\indent The genome consists of 11 genes, each of which has its 
condition (left-hand side) and its expression (right-hand side). Genes 1 to
4 control cell division into the different types that are needed, as
well as the growth of axons and dendrites.
The first gene is only expressed by the initial cell, and only
if no internal protein \texttt{\small ip} is
present (the sequence \small\texttt{NNY(ip) SUP(cpt)}\normalsize).
In addition, gene 1 is 
\begin{figure*}[!tbp]
\begin{center}
\epsfig{file=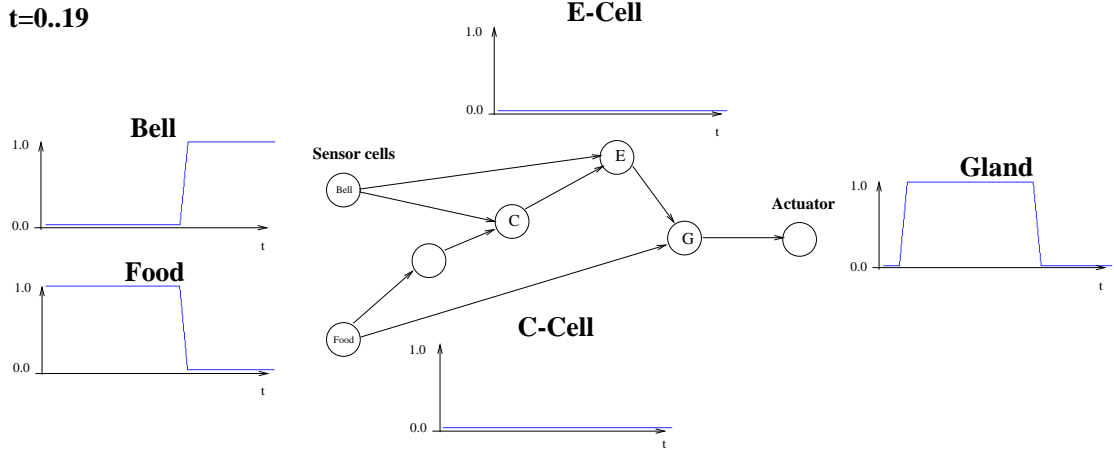,angle=-90,width=14.7cm}
\caption{{\small\label{fig:classicalCond1} First, only food is
    present. This triggers the gland because of the hardwired reflex,
    while the ``C''-cell and ``E''-cell remain inactivated. Later, only
    the bell signal is present. Due to the fact that the network is
    not yet conditioned, none of the cells becomes active as a result.}}
\end{center}
\end{figure*}
\begin{figure*}[!tbp]
\begin{center}
\epsfig{file=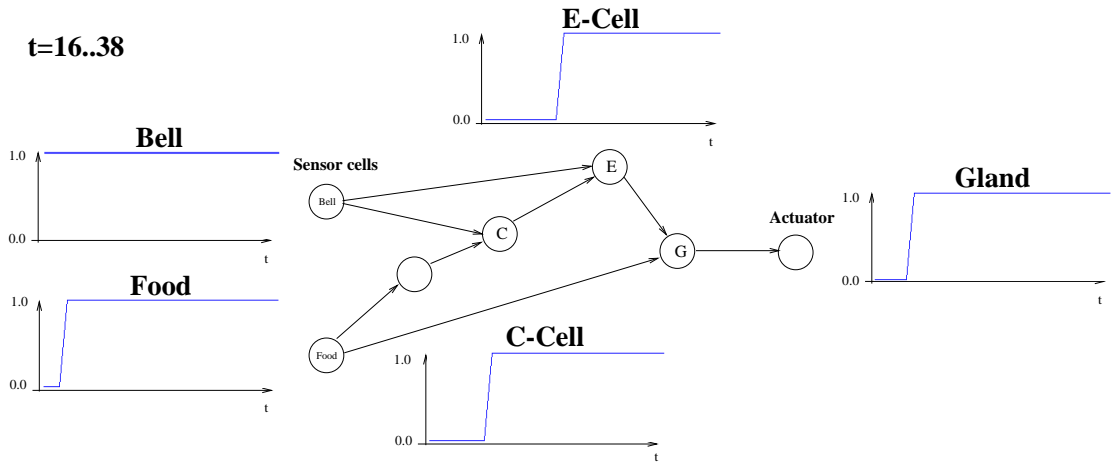, angle=-90,width=14.7cm}
\caption{{\small \label{fig:classicalCond2} Both sensors, food and
    sound, are stimulated. The ANN becomes conditioned (``C''-cell)
    and the gland is triggered due to the presence of food.}}
\end{center}
\end{figure*}
\begin{figure*}[!tbp]
\begin{center}
\epsfig{file=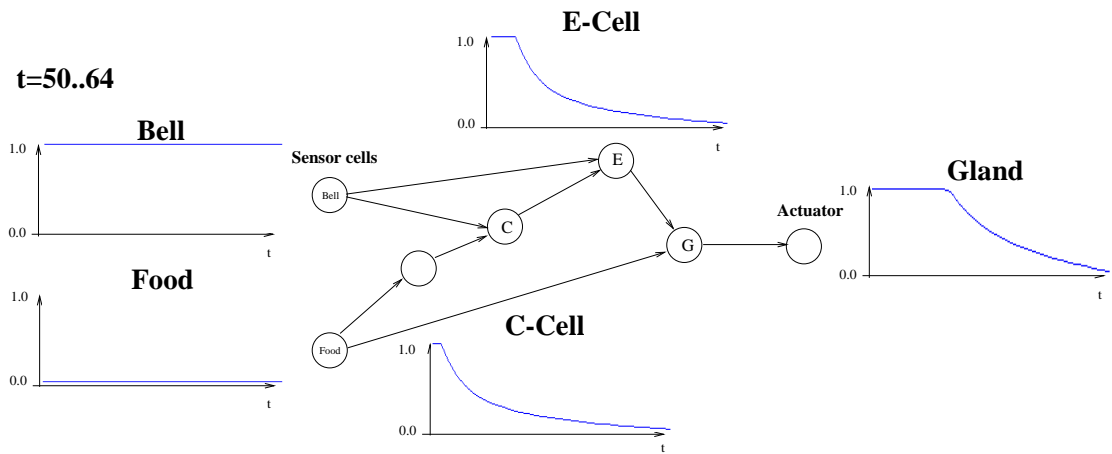, angle=-90,width=14.7cm}
\caption{{\small\label{fig:classicalCond3} Being in the conditioned
state, the food sensor suddenly becomes deactivated while the bell keeps on
ringing. Thus, the activation of the ``C''-cell becomes weaker. This
implies a decrease of activation of the ``E''-cell which finally
results in a decline of gland activity.}}
\end{center}
\end{figure*}
only expressed if it senses nonzero
concentrations of cell-type protein \texttt{\small spt0} 
(\small\texttt{ANY(spt1) ANY(spt0)}\normalsize) which is emitted by one of the sensor cells and has diffused. 
Under these circumstances the initial cell will divide and produce offspring
of type \texttt{\small acpt0} and \texttt{\small acpt2}
(\small\texttt{SPL(acpt2) SPL(acpt0)}\normalsize), 
grow a dendrite that follows the gradient of the sensor protein \texttt{\small spt0} (\small\texttt{GDR(spt0)}\normalsize), and produce
the internal protein \texttt{\small ip}. Once \texttt{\small ip}
is produced, this will continue to happen (gene 5) which 
prevents that gene 1 can ever be turned on again. 
Genes 2 to 4 work just as gene 1, but for other cell types.
While gene 5 takes care of the hardwired-reflex stimulation,
genes 10 and 11 control the conditioning (expressed
only by cell-type \texttt{\small cpt}). If
food is present and the bell rings, gene 10 is expressed. It
produces certain amounts of external protein
\texttt{\small ep}. The concentration of \texttt{\small ep} influences
cell stimulation (gene 11 exhibits stimulation via
\texttt{\small EXT}). Due to diffusion,
\texttt{\small ep} diminishes over time, so the
conditioning decreases accordingly.
As pathways to the ``C'' cell are not equally long, a production
cascade of internal proteins in genes 8, 9 and 10 is necessary that
delays the input of the acoustical sensor.
The behavior of the resulting phenotype network is documented
in Figures \ref{fig:classicalCond1}, \ref{fig:classicalCond2} and 
\ref{fig:classicalCond3}.

\section{WWW-based Genetic Algorithm: Community of Artificial Organisms}

While it is not difficult to write genomes which lead to simple
networks with desired characteristics, one of the main features of the
system is its evolvability. Certainly, the search space for such
genomes is immense, and it is unreasonable to hope that interesting
genomes can be found without massive parallelism. Rather than choosing
to implement this system on supercomputers, we opted to
allow users on the Internet to donate their CPU time by participating
in a global evolutionary experiment.


Using Sun's Java$^{\rm TM}$ technology, we developed an asynchronous,
distributed GA system which allows a massive parallel search for new
genomes based on evolutionary 
principles \cite{astorthesis,astorpaper}. 
It consists mainly of a central server application and clients, each
of which hosting {\em one} individual of the current GA 
population.
\begin{figure*}[!tbp]
\begin{center}
\includegraphics[angle=-90, width=8cm]{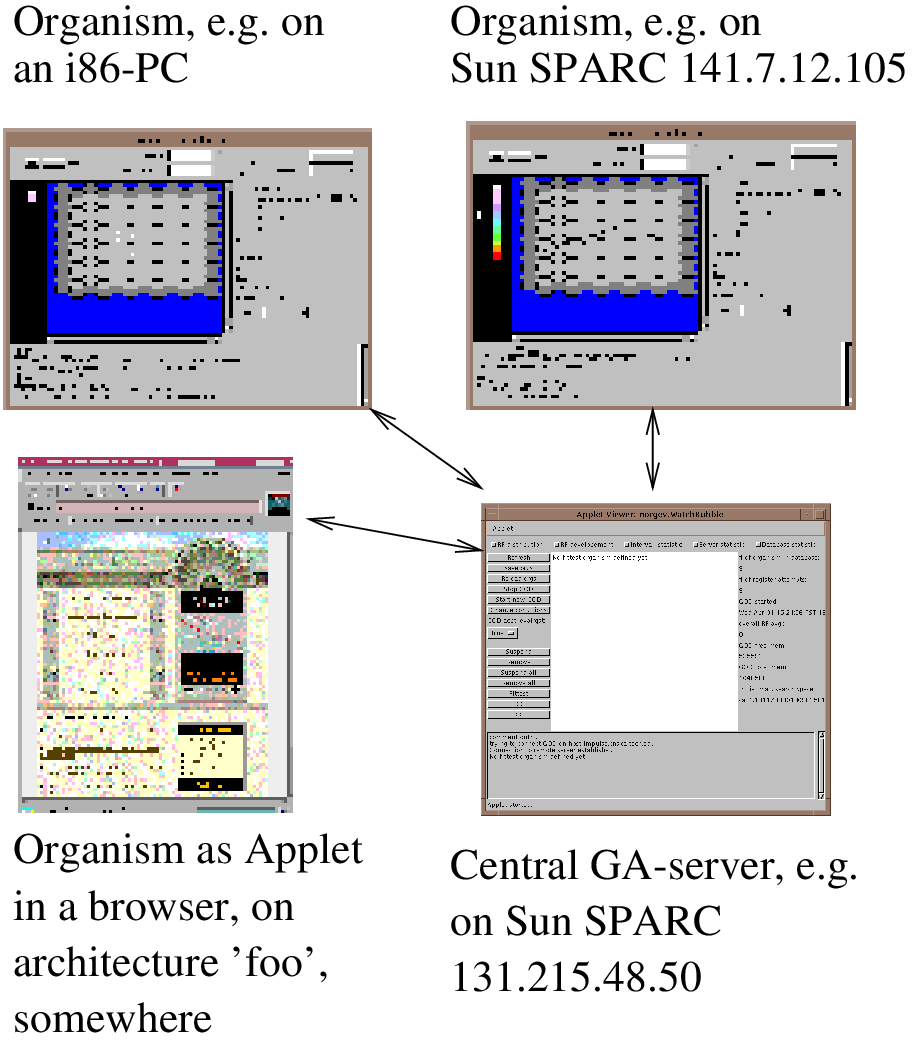}
\includegraphics[angle=-90, width=8cm]{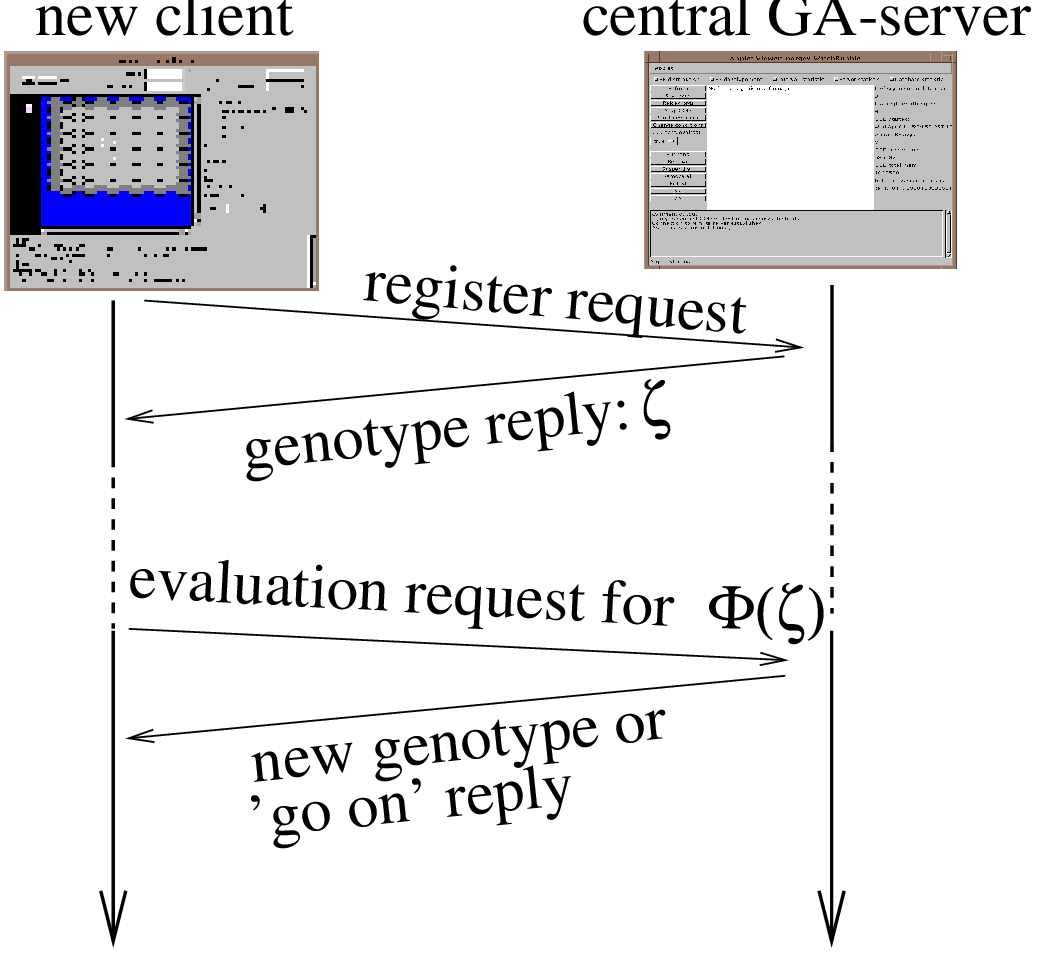}
\end{center}
\caption{\label{fig:distrfig}\label{fig:commfig}Left: genotype 
evaluation in clients of different architecture, using TCP/IP for
communication with the asynchronous genetic algorithm. Right:
communication between the GA-server and a client hosting an organism.}
\end{figure*}

As Java is supposed to be platform independent, clients can be started from
every computer for which an accurate Java 1.1 virtual machine or
browser exists (Figure \ref{fig:distrfig}).
The clients are hybrids, which means that they can be started as Java
Applet by choosing the html page from our WWW-server, or with the
help of a bootloader program which dynamically downloads the client
and starts it as a Java application (no browser necessary).

A client automatically sends a request-to-register to the central
GA-server after it was started, and receives from the server a 
genotype $\zeta$. The client
then starts up a simulation as described in Section
\ref{sec_simulation}. After a
certain number of simulation cycles, it sends the genotype's fitness
$\Phi(\zeta)$ (average reinforcement signal during simulation) to the
server.
By comparing $\Phi(\zeta)$ to the fitness of other genotypes in the
database, the server decides if it is worth to keep this genotype or if
the client should be assigned a new one. If the
server has to send a new genotype, it either takes a suspended one out
of its database, or {\em constructs} one through the processes
of recombination and/or mutation from genotypes of known
fitness already present in the population (Figure \ref{fig:commfig}).
Fitter genotypes are more likely to be selected for recombination
and/or asexual copying then genotypes of lower fitness. This leads to
an increase of the average fitness over time.

\section{Conclusion}

We introduced a developmental and behavioral model based on artificial
gene expression which shares key properties with natural neural
development. Within this model we succeeded to construct
simple systems with properties which 
are believed to be essential~\cite{kandel1,koch1} for
higher self-organizing information processing systems, such as deterministic
structure development, self-limiting growth, growth
following diffusion gradients, computation of logical functions, pacemaker behavior and simple adaptation (sensitization, habituation, associative classical
conditioning) \cite{astorthesis,astorpaper}.
Furthermore, we showed how an evolutionary search for
genomes coding for information-pro\-cess\-ing network structures 
can be distributed in a platform-independent manner such that
the unused CPU power of the Internet can be tapped to search for ANNs
that reduce the gap between the abstract models and neurophysiology. 


\end{document}